\documentclass[12pt]{article}

\usepackage{arxiv}

\usepackage[utf8]{inputenc} 
\usepackage[T1]{fontenc}    
\usepackage{url}            
\usepackage{booktabs}       
\usepackage{amsfonts}       
\usepackage{nicefrac}       
\usepackage{microtype}      
\usepackage{lipsum}
\usepackage{graphicx}

\usepackage{xcolor}

\usepackage[backend=biber,sorting=none]{biblatex}
\usepackage{array}

\usepackage{amsfonts}       
\usepackage{mathtools}
\usepackage{enumitem}
\usepackage{bm}
\usepackage{amsmath}
\usepackage{amsthm}
\usepackage{amssymb}
\usepackage{pifont}
\newcommand{\eq}{\begin{equation*}}
\newcommand{\en}{\end{equation*}}
\newcommand{\eqa}{\begin{eqnarray*}}
\newcommand{\ena}{\end{eqnarray*}}
\newcommand{\eqn}{\begin{equation}}
\newcommand{\enn}{\end{equation}}
\newcommand{\be}{\begin{equation}}
\newcommand{\ee}{\end{equation}}
\newcommand{\eqan}{\begin{eqnarray}}
\newcommand{\enan}{\end{eqnarray}}

\usepackage[colorlinks=true, allcolors=blue]{hyperref}
\usepackage[noabbrev,capitalize,nameinlink]{cleveref}

\usepackage{amsfonts}
\newcommand{\pmat}{\begin{pmatrix}}
\newcommand{\pman}{\end{pmatrix}}

\usepackage{authblk}

\title{Super-Efficient Exact Hamiltonian Monte Carlo \\
for the von Mises Distribution
}

 \author{
 Ari Pakman \\
IEM Department, Ben-Gurion University of the Negev, Beer Sheva, Israel 
 }
\begin{document}
\maketitle
\begin{abstract}
Markov Chain Monte Carlo algorithms, the method 
of choice to sample from generic high-dimensional distributions, 
are rarely used for continuous one-dimensional distributions, 
for which  more effective approaches are usually available (e.g. rejection sampling). 
In this work we present a counter-example to this conventional wisdom
for the von Mises distribution,
a  maximum-entropy distribution over the circle. 
We show that Hamiltonian Monte Carlo with Laplacian momentum has exactly solvable equations of motion and, with an appropriate travel time, the Markov chain has negative autocorrelation at odd lags for odd observables and  yields a relative effective sample size bigger than~one.  
\end{abstract}


\section{Introduction}
The von Mises distribution~\cite{vonMises1918} has density
\eqan 
\pi(x) = \frac{\exp(\kappa \cos(x-\nu))}{2 \pi I_0(\kappa)}   \qquad x\in [-\pi, +\pi] \,, 
\label{eq:vMises}
\enan 
where $\kappa > 0$ and $I_0(\kappa)$ is the modified Bessel function of the first kind of order 0. It is the maximum entropy distribution in the circle with fixed  real and imaginary parts of the first circular moment. It is important in statistical models of circular data, and arises in many areas of science, including astronomy, biology, physics, earth science and meteorology~\cite{fisher1995statistical,jammalamadaka2001topics,pewsey2013circular}.

Since inverting the cumulative distribution of (\ref{eq:vMises}) is intractable, several algorithms  to generate pseudo-random samples from 
(\ref{eq:vMises}) have been proposed~\cite{best1979efficient,dagpunar1990sampling,marsaglia1998monty},
the most popular being the accept-reject method of Best and Fisher~\cite{best1979efficient}. In this work we present a more efficient approach based on the Hamiltonian Monte Carlo (HMC) method~\cite{duane1987hybrid}.

\section{Review of Hamiltonian Monte Carlo}
For completeness, we present here the HMC method specialized to one-dimensional distributions. 
For  a thorough review in higher dimensional spaces see~\cite{neal2010mcmc}.  
The idea is to create a Markov chain with equilibrium distribution $\pi(x)$  out of the endpoints of successive trajectories of a classical mechanical particle with potential energy~$U(x) = -\log \pi(x)$. The sample space is augmented  with a momentum variable~$p \in \mathbb{R}$, 
with kinetic energy~$K(p)$ and distribution $\mu(p) =  c e^{-K(p)}$, where~$c$ is a normalization constant and  
$K(p)$ can be chosen freely as long  as~$\mu(p)$ is normalizable and $K(p) = K(-p)$. 
Let us define the Hamiltonian function
\begin{equation}
    H(x,p) = U(x) + K(p) = -\log \pi(x) -\log \mu(p)  + \text{const.}
\label{eq:H}
\end{equation}
The HMC Markov chain preserves the 
joint distribution $\tilde{\pi}(x, p) = \pi(x)\mu(p) = e^{-H(x,p)}$ and is defined as follows. 
Let us denote successive  $x$-states in the chain as $x_0, x_1 \ldots$
The transition~$x_k \rightarrow x_{k+1}$, 
involves two steps:
\begin{itemize}

    \item[(a)] Sample $p\sim e^{-K(p)}$

    \item[(b)] 
    Assuming initial conditions   $x(0) = x_k, p(0) = p$, 
    evolve  $x(t), p(t)$  for a  time~T with  the Hamiltonian  equations of motion 
    \begin{equation}
\label{eq:EOM}    
\dot{x}(t) = \frac{\partial H}{\partial p} 
\qquad \qquad
\dot{p}(t) =  - \frac{\partial H}{\partial x},         
    \end{equation} 
    and let $x^{(k+1)}= x(T)$.
\end{itemize}

The travel time $T$ is a free hyperparameter that affects the mixing of the chain. 
Step $(a)$ clearly preserves the 
distribution~$\tilde{\pi} (x,p)$. 
To analyze step~(b), it is convenient to define the  map $F(x,p) = (x(T), -p(T))$.
Thus step~(b) corresponds to 
a deterministic transition kernel
\begin{equation}
R(x', p'|x, p) \equiv \delta((x', p') - F(x,p)  ) \,.
\label{eq:HMC_kernel}
\end{equation}
By symmetry, if the  trajectory starts at $(x', p')$ it 
will be mapped to  $(x, p)$ by $F$, and thus the map satisfies
$F^2(x,p) = (x,p)$.  Importantly, Hamiltonian dynamics (\ref{eq:EOM}) imply two well-known  conservation laws~\cite{goldstein2001}: 
\begin{itemize}
    \item 
    {\bf Conservation of Energy:} 
    \[ H(x,p)  = H(x',-p')\]
    \item 
    {\bf Conservation of Volume:} 
    \\
    \begin{center}
    $dx dp = dx'dp'$ or equivalently 
$|\nabla_{x,p}F(x,p)| =1$.        
    \end{center}

\end{itemize}
 Using both conservation laws, it is immediate to show that the Markov kernel~(\ref{eq:HMC_kernel}) satisfies detailed balance, 
\begin{align}
\tilde{\pi} (x,p) R(x', p'|x, p)  &= e^{-H(x,p)}\delta((x', p') - F(x,p)  ) 
\nonumber
\\
& = e^{-H(x', \, p')}\delta((x', p') - F(x,p)  ) 
 && \text{\small Energy conservation and $K(p)=K(-p)$}
\nonumber 
\\
&  = e^{-H(x', \,p')}\delta((x,p) -F^{-1}(x', p')    )|\nabla_{x,p}F|^{-1} 
 && \text{\small  
 Change of variables}
\nonumber 
\\
&   = \tilde{\pi}(x', p') \delta((x,p) - F(x', p')  )   
&& 
\text{\small   $F^2=1$ and  $|\nabla_{x,p}F| =1$.
}
\nonumber 
\\
&   = \tilde{\pi}(x',p')  R(x, p|x', p') \,. && \text{\small }
\nonumber 
\end{align}
Thus, under standard conditions~\cite{tweedie1975sufficient}, 
the Markov chain defined by  (a)-(b) 
converges to $\tilde{\pi}(x,p) $.

\section{Exact solutions for the von Mises distribution}
For most distributions $\pi(x)$ of interest, the equations (\ref{eq:EOM}) can only be solved approximately via numerical integration, and the final values $x(T), p(T)$ become a proposal for a Metropolis-Hastings accept-reject step. Known distributions~$p(x)$ for which the equations
(\ref{eq:EOM}) are exactly solvable are 
 locally trivial, such as Gaussian, with $U(x)\varpropto x^2$, or 
 Exponential, with $U(x) \varpropto x$, although they can be 
 globally non-trivial~\cite{pakman2013auxiliary,pakman2014exact}. 
Here we show that also for the von Mises distribution~(\ref{eq:vMises}) an exact solution is  possible. 

The key is to chose a Laplace distribution for the momentum, 
$p \sim \mu(p)= \frac12 e^{-|p|}$, which yields the Hamiltonian (\ref{eq:H})  
\begin{align}
H &=   -\kappa \cos(x) + |p| +   \text{const.}
\label{eq:H_Laplace}
\end{align}
Laplace momentum is rarely used for Hamiltonian Monte Carlo, with a notable exception being~\cite{nishimura2020discontinuous}, but in our case it is crucial 
since it yields equations of motion (\ref{eq:EOM})
\begin{align}
\dot{x}(t) &  = s, \qquad \qquad  \text{where} \,\, s= \text{sign}(p) , 
\\
\dot{p}(t) &=  - \kappa \sin(x) ,
\end{align}
which are solved exactly as
\begin{align}
x(t) &= x_0 + st, 
\label{eq:sol1}
\\
p(t) &=  p_0 - \kappa s \cos(x_0)  + \kappa s \cos(x_0 + st). 
\label{eq:sol2}
\end{align}
Here we denoted  $x_0 = x(0), p_0=p(0)$. 
This solution is valid until $p(t)$ crosses zero and changes its sign,
which can happen if the initial values $(x_0, p_0)$ are such that 
\begin{align}
   \cos(x_0)  - \frac{|p_0|}{\kappa} \geq -1 \,.
\label{eq:m_zero_condition}
\end{align}
If (\ref{eq:m_zero_condition}) holds, the time $t_{z,1}>0$ such that $p(t_{z,1})=0$ can be found analytically from (\ref{eq:sol2}) in terms of the $\arccos$ function. After evolving~(\ref{eq:sol1})-(\ref{eq:sol2}) until time $t_{z,1}$, we let $p_0 \leftarrow 0, s \leftarrow -s, x_0 \leftarrow x(t_{z,1})$ and the trajectory continues according 
to~(\ref{eq:sol1})-(\ref{eq:sol2}) until $p(t_{z,2})=0$ again or until the accumulated elapsed time reaches  $T$. Repeating this procedure, the final position in each iteration is 
\begin{equation}
    x_{i+1} = x_{i} - s_0  \sum_{i=1}^{q+1} (-1)^i t_{z,i} \qquad \text{where } s_0 = \text{sign}(p(0))
\end{equation}
where $T = \sum_{i=1}^{q+1} t_{z,i}$, $q$ is the number of times the momentum
crossed $p=0$, and each $t_{z,i}$ is a time interval at the end of which either $p=0$ or the total  time $T$ has been reached.

\begin{figure*}[t!]
\begin{center}
\includegraphics[width=1\textwidth]{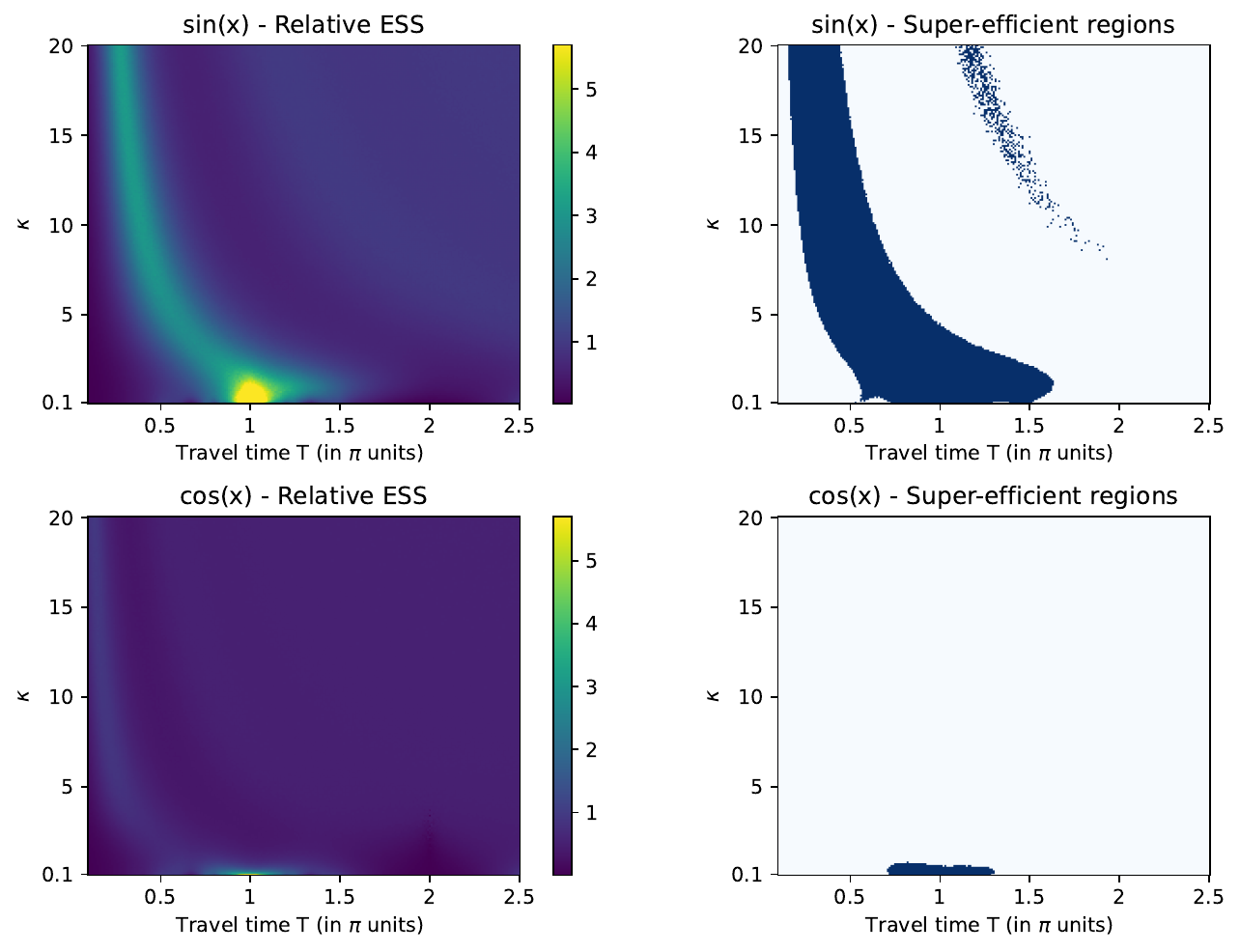}
\end{center}
\caption{ {\bf Left panels:}  Relative Effective Sample Size (RESS), defined in~(\ref{eq:RESS}), for the chains of $\sin(x)$ and $\cos(x)$, estimated from  $N=10^5$ iterations of the HMC Markov chain, for a range of values of $(\kappa, T)$.
{\bf Right panels:} Values of $(\kappa, T)$ where RESS $>1$ are indicated in blue and correspond to super-efficient sampling.
}
\label{fig:ESS}
\end{figure*}

\begin{figure*}[t!]
\begin{center}
\includegraphics[width=1\textwidth]{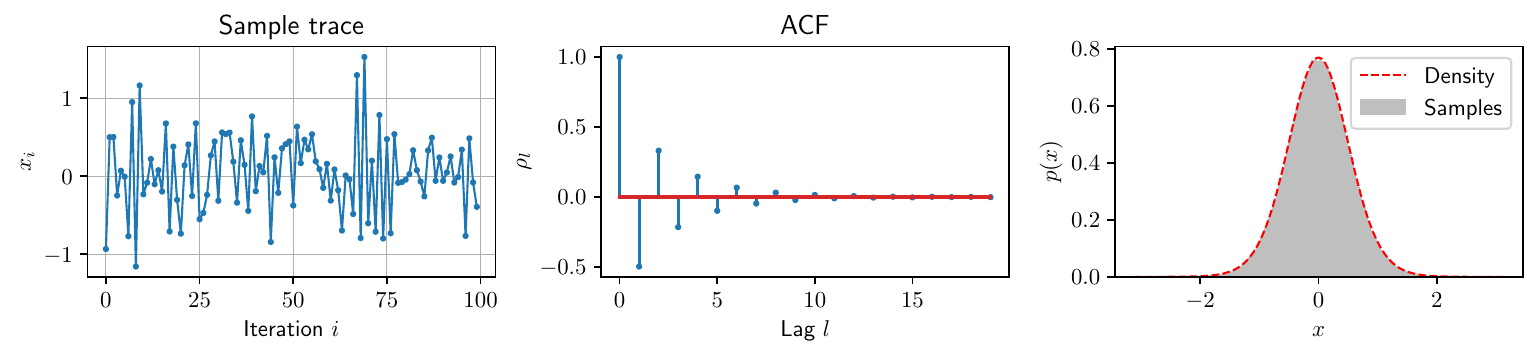}
\end{center}
\caption{The HMC algorithm applied to the von Mises distribution  with $\kappa = 4, \nu = 0$. The travel time was $T=2.32$, which yields the highest relative ESS according to the results in~\cref{fig:optimal} {\it (Upper right)}. {\bf Left:} sample trace for 100 iterations. Note the tendency of successive samples to change sign. 
{\bf Middle:} Auto-correlation function of $\sin(x)$ with alternating signs. The same pattern occurs all values of $\kappa$. 
{\bf Right:} Histogram of samples vs. analytic density.
}
\label{fig:k4_samples}
\end{figure*}

\section{Numerical results}
To evaluate the method, we simulated the Markov chain for a range of $\kappa \in [0.1, 20]$ and travel times~$T \in [0, 2.5 \pi]$.  Without loss 
of generality, we assumed $\nu=0$. 
We chose the observables  $\sin(x_i)$  and $\cos(x_i)$, instead of~$x_i$, to avoid artifacts related to the periodicity $x_i = x_i + 2 \pi$. We measure the efficiency via the Relative Effective Sample Size~\cite{liu2001monte},\footnote{For the estimation we used the \texttt{arviz} python package~\cite{kumar2019arviz}.}
\begin{equation}
    \text{RESS} = \frac{1}{1 + 2\sum_{l=1}^{\infty} \rho_l},
\label{eq:RESS}
\end{equation}
where $\rho_l$ is the autocorrelation at lag $l$ 
of the chain $\{\sin(x_i)\}_{i=1}^{\infty}$ or $\{\cos(x_i)\}_{i=1}^{\infty}$. 
The RESS estimates are shown in~\cref{fig:ESS}.
 Note that for  $\sin(x)$, for every $\kappa$ there 
 exists  a segment of travel times $T$ for which RESS~>~1. This corresponds to 
super-efficient  sampling, where the variance of the estimate of the mean of $\sin(x)$ is lower than that obtained using uncorrelated i.i.d. samples. The reason is that at such values of $T$ the Markov chain is antithetic, having negative autocorrelations at odd lags. \cref{fig:k4_samples} {\it (Middle)} shows this in detail for~$\kappa =4$,
and similar results holds for other values of $\kappa$. 
Of course, for even observables such as $\cos(x)$ 
the chain loses its antithetical properties and we get RESS~<~1 for most parameter values, as illustrated in~\cref{fig:ESS}.

\begin{figure*}[b!]
\begin{center}
\includegraphics[width=\textwidth]{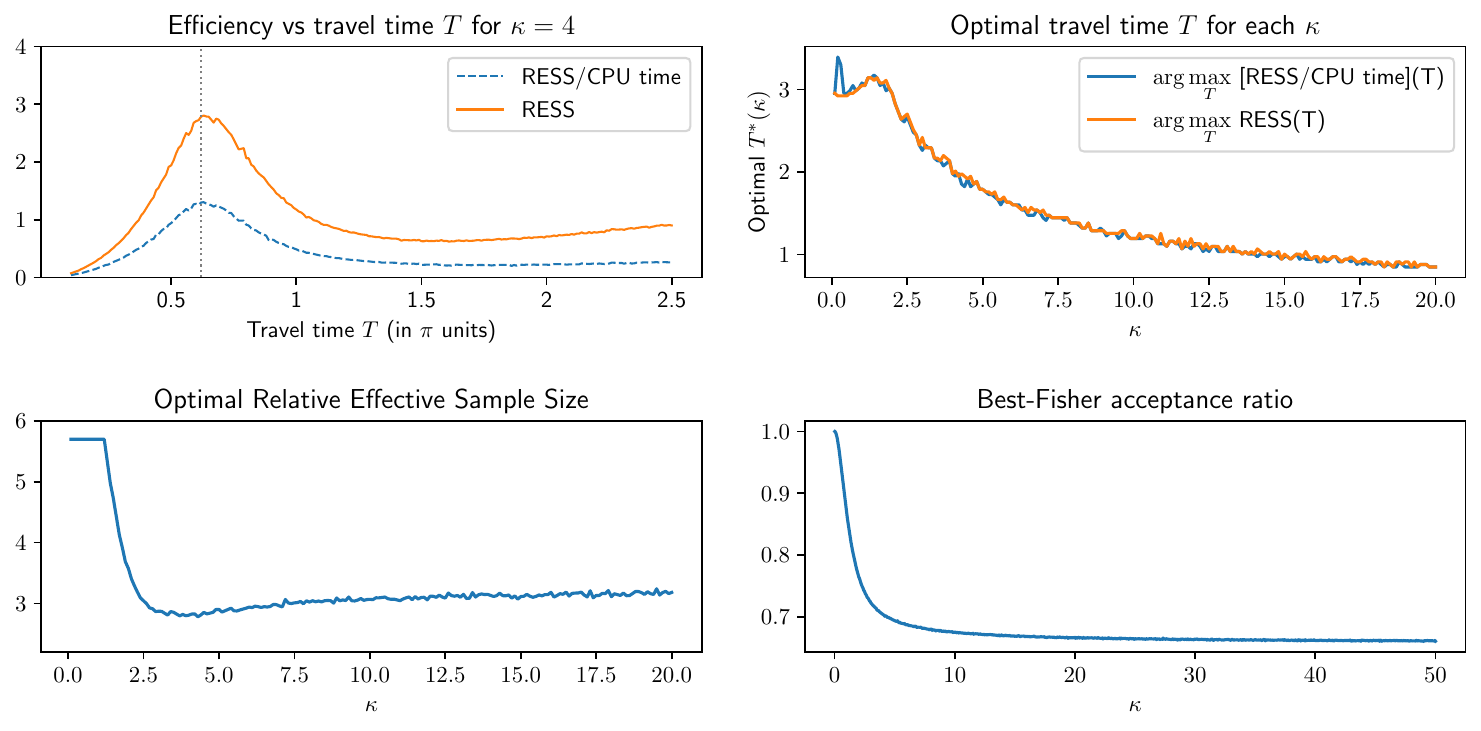}
\end{center}
\caption{ 
{\bf Upper left:} Efficiency of the $\sin(x)$ chain as a function of $T$ for the $\kappa=4$
example of~\cref{fig:k4_samples}. 
{\bf Upper right:} Optimal travel time~$T^*(\kappa)$ for the $\sin(x)$ observable, obtained from the results of~\cref{fig:ESS}. Note the negligible difference between optimizing RESS or RESS/CPU time.
{\bf Lower left:} Optimal $\sin(x)$ RESS using $T^*(\kappa)$ as a function of $\kappa$. 
{\bf Lower right:} Efficiency of the accept-reject method of~\cite{best1979efficient}. 
}
\label{fig:optimal}
\end{figure*}

\cref{fig:optimal} {\it (Upper right)} shows, for each $\kappa$,  the optimal travel time $T^*(\kappa)$, 
defined as the value of $T$ for which the $\sin(x)$ RESS is maximal. In a practical implementation, 
the function $T^*(\kappa)$ can be hard-coded into the algorithm. 
We leave a theoretical derivation of $T^*(\kappa)$ for future work. 
Note also, \cref{fig:optimal} {\it (Lower left)},  that the at the optimal $T^*(\kappa)$  we have $\text{RESS} \simeq 3$ for a broad range of $\kappa$. This stands in contrast with the accept-reject method of~\cite{best1979efficient}, whose acceptance rate 
 shown in \cref{fig:optimal} {\it (Lower right)}, 
diminishes towards $\simeq .65$ with growing $\kappa$.

\begin{figure*}[h!]
\begin{center}
\includegraphics[width=.49\textwidth]{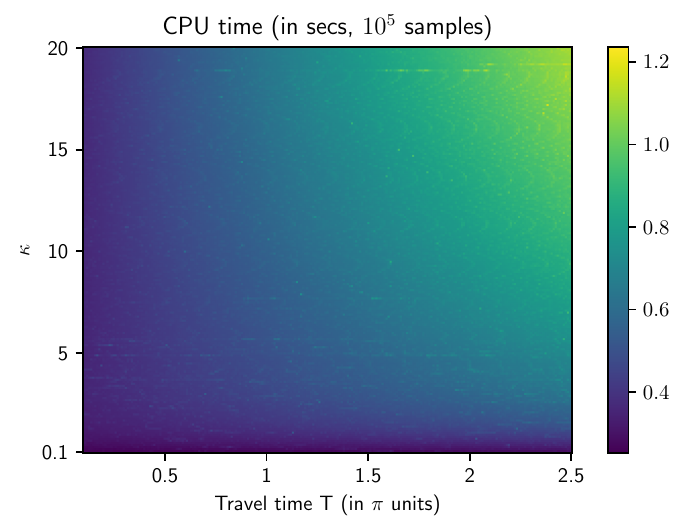}
\end{center}
\caption{CPU time  for $N=10^5$ iterations of the HMC Markov chain for the same range of $\kappa$ and travel times $T$ presented in ~\cref{fig:ESS}}
\label{fig:CPU}
\end{figure*}



\section{Related works}
The possibility of super-efficient sampling using Markov chains was noticed already by Peskun~\cite{peskun1973optimum}, 
who provided, in finite dimensions,  
a necessary and sufficient condition on the spectral properties of a Markov transition matrix,  such that expectations estimated from Markov chain samples have lower asymptotic variance than using independent samples. 

The Laplace momentum we used in  (\ref{eq:H_Laplace}) implies that the velocity of the particle is restricted to $s \in \{ \pm 1\}$. The resulting dynamics are reminiscent of the Zig-Zag sampler~\cite{peters2012rejection}, 
whose one-dimensional version was studied in detail in~\cite{monmarche2014hypocoercive, fontbona2016long, bierkens2017piecewise}. Similar to HMC, 
the Zig-Zag sampler augments the sampling space a velocity, but unlike HMC, it does not satisfy detailed balance. 
For Zig-Zag samplers in generic dimensions, a different notion of super-efficiency was shown to hold in~\cite{bierkens2019zig}. In this case,
independent samples of the parameters of a Bayesian posterior distribution based on $n$ data points can be obtained 
with a computational cost of~$O(1)$ per sample, instead of $O(n)$ (after a one-time initial $O(n)$ cost), by using an appropriate control variate.

\section{Conclusions}
In this work we have presented a simple Markov chain based on exactly solvable HMC dynamics to sample from the von Mises distribution. 
Possible extensions include  generalizations to higher dimensional von Mises distributions, such as those studied in \cite{navarro2017multivariate,cohen2024mises}. 
More generally, the use of Laplace momenta in HMC algorithms 
has been barely explored, and we hope that this works opens the way to more models were Laplace momenta could be of benefit. 

\section*{Acknowledgments}
This research was supported by the Israel Science Foundation (grant No. 1138/23).

\clearpage 
\printbibliography
\end{document}